# Comparative analysis of specific heat of YNi$_2$B$_2$C using nodal and two-gap models

C. L. Huang,[1] J. –Y. Lin,[2] C. P. Sun,[1] T. K. Lee,[3] J. D. Kim,[4] E. M. Choi,[4] S. I. Lee[4,5]
and H. D. Yang[1,3*]

[1]*Department of Physics, National Sun Yat-Sen University, Kaohsiung 804, Taiwan*
[2]*Institutre of Physics, National Chiao-Tung University, Hsinchu 300, Taiwan*
[3]*Institute of Physics, Academia Sinica, Nankang 11592, Taiwan*
[4]*National Creative Research Initiative Center for Superconductivity and Department of Physics, Pohang University of Science and Technology, Pohang, 794-784, Republic of Korea*
[5]*Quantum Material Laboratory, Korea Basic Science Institute, Daejeon, 305-333 Korea*

The magnetic field dependence of low temperature specific heat in YNi$_2$B$_2$C was measured and analyzed using various pairing order parameters. At zero magnetic field, the two-gap model which has been successfully applied to MgB$_2$ and the point-node model, appear to describe the superconducting gap function of YNi$_2$B$_2$C better than other models based on the isotropic $s$-wave, the $d$-wave line nodes, or the $s+g$ wave. The two energy gaps, $\Delta_L$=2.67 meV and $\Delta_S$=1.19 meV are obtained. The observed nonlinear field dependence of electronic specific heat coefficient, $\gamma(H) \sim H^{0.47}$, is quantitatively close to $\gamma(H) \sim H^{0.5}$ expected for nodal superconductivity or can be qualitatively explained using two-gap scenario. Furthermore, the positive curvature in $H_{c2}(T)$ near $T_c$ is qualitatively similar to that in the other two-gap superconductor MgB$_2$.

Rare-earth nickel borocarbides RNi$_2$B$_2$C (R=Dy, Ho, Er, Tm, Lu, and Y) have been among the most studied superconductors during the past decade. The general interest centers around their many intriguing physical properties, such as the relatively high superconducting transition temperature $T_c$~15 K, the coexistence of superconductivity and long-range magnetic order, and the reentrant superconductivity.[1,2] After having extensive theoretical and experimental works,[3] it remains unclear whether they are conventional or exotic superconductors.[4] Particular attention has been aimed at determining the superconducting order parameter, which is thought to be essential in establishing a microscopic model of its superconductivity. At an early stage, the order parameter in YNi$_2$B$_2$C was considered to be a conventional isotropic $s$-wave pairing and mediated by conventional electron-phonon interactions.[5-8] However, recent thermal and spectroscopic experiments indicate the high anisotropy[9-19] and $d$-wave pairing with line nodes.[20-22] Furthermore, angular dependence of the thermal conductivity,[12-14] specific heat,[15,16,23] point-contact tunneling[17] and ultrasonic attenuation[18] all seem to provide evidence of point nodes in the superconducting gap function. Accordingly, a hybrid $s + g$ pairing symmetry has been proposed for this material.[24-26] On the other hand, the multiband superconductivity in YNi$_2$B$_2$C has recently been proposed based on the observations from the temperature dependence of the upper critical field[27,28] and the directional point-contact spectroscopy.[29,30]

In this Letter, we present specific-heat data of YNi$_2$B$_2$C. By fitting them to various superconductivity models, it is found that two-gap model, which has been successfully applied to MgB$_2$, best describes the gap function of superconducting YNi$_2$B$_2$C.

The single crystal YNi$_2$B$_2$C used for the present study was grown by the high temperature Ni$_2$B flux method. Details of preparation have been published elsewhere.[31] The low temperature specific heat $C(T,H)$ was measured with a $^3$He heat-pulsed thermal relaxation calorimeter[32,33] in the temperature range from 0.6 to 20 K and under magnetic fields up to 8 T. The magnetic field was applied in the direction perpendicular to the c-axis of the crystal.

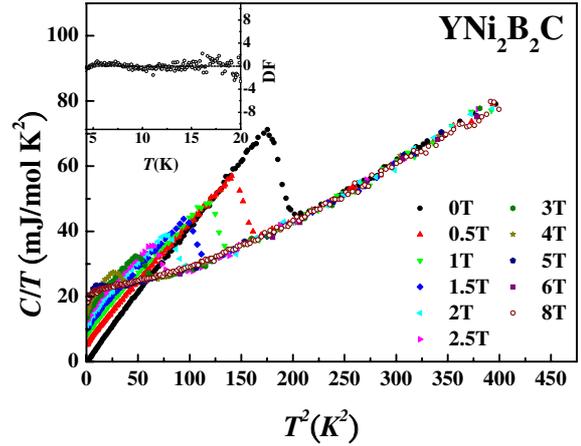

**FIG. 1.** Temperature ($T$) dependence of specific heat ($C$) plotted as $C/T$ vs. $T^2$ at various magnetic fields for YNi$_2$B$_2$C. The inset shows the deviation of the fit of 8 T data to $C_n(T) = AT^{-2} + \gamma_n T + \beta T^3 + \alpha T^5$.

Figure 1 shows the results of calorimetric measurements $C(T,H)$ of YNi$_2$B$_2$C as $C/T$ vs. $T^2$ with $H$ varying between 0 and 8 T. Occurrence of superconductivity becomes evident through a specific-heat jump at the temperature $T_c$ for a given field. At zero field, the extrapolation of the data below $T_c$ to 0 K points to a zero intercept, indicating a full superconducting volume fraction and confirming the good sample quality. The $T_c$ value of 13.77 K as determined from the midpoint of specific-heat jump is consistent with that obtained from resistivity and magnetization (not shown). The normal-state specific heat at zero magnetic field can be simply described by

$$C_n(T) = \gamma_n T + C_{lattice}(T), \quad (1)$$

where $\gamma_n T$ is the electronic term due to free charge carriers and $C_{lattice}(T) = \beta T^3 + \alpha T^5$ represents the phonon contribution which is assumed to be independent of the magnetic field. In order to achieve an optimal normal-state fitting to Eq. (1) for $H=8$ T, the data between 3.5 and 20 K are used but with an additional $T^{-2}$ hyperfine-contribution term, which is thought to be due to a very low concentration of paramagnetic centers. The best fitting parameters, $\gamma_n = 19.74 \pm 0.27$ (mJ/mol K$^2$), $\beta = 0.077 \pm 0.003$ (mJ/mol K$^4$) corresponding to $\Theta_D = 533 \pm 7$ K and $\alpha = 0.00018 \pm 0.00001$ (mJ/mol K$^6$), are fairly consistent with those previously reported.[7,9] In fact, these parameters are justified by the entropy balance, entropy balance $S = \int_0^{T_c} \frac{\delta C}{T} dT$ at the second-order superconducting-normal phase transition supporting the validity of the fitting above, where $\delta C(T) = C(T, H=0) - C_n(T)$.

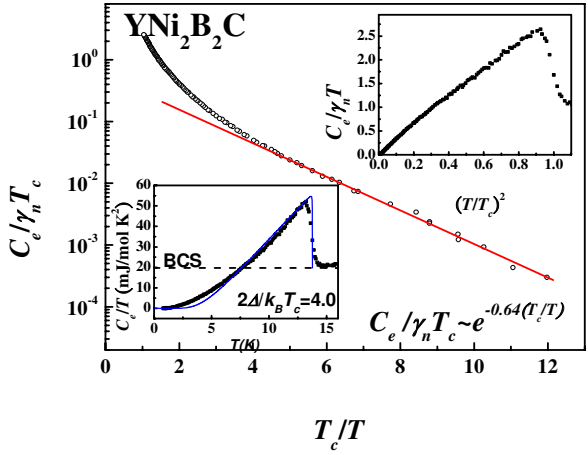

**FIG. 2.** A logarithmic plot of $C_e/\gamma_n T_c$ vs. $T_c/T$ for YNi$_2$B$_2$C in its superconducting state. The solid line is the linear fit to the data for $T_c/T$ between 4 ($T$=3.7 K) and 12 ($T$=1.1K). The left-side inset shows the data fitted to the BCS model. In the right-side inset, the almost linear relation between $C_e/\gamma_n T$ and $(T/T_c)^2$ indicates a nearly $T^3$-dependence of $C_e$.

The electronic specific heat in the superconducting state ($T < T_c$) is given by $C_e(T) = C(T) - C_{lattice}(T)$. The logarithmic plot of $C_e(T)/\gamma_n T_c$ vs. $T_c/T$ in Fig. 2 shows a linear fitting of the data between $T_c/T = 4$ ($T \sim 3.7$ K) and 12 ($T \sim 1.1$ K) following the relation $C_e(T)/\gamma_n T_c \sim \exp[(-a T_c/T)]$ with $a=0.64$. The parameter $a$ is related to the superconducting energy gap. Realizing that the BCS theory predicts $C_e(T)/\gamma_n T_c \sim \exp[(-1.44 T_c/T)]$ in the weak coupling limit, the fitted value of $a$ is too small to support such a relatively high $T_c \sim 13.77$ K for a single fully gapped BCS-like superconductor. Furthermore, data points in Fig. 2 deviate notably from linearity at temperatures higher than around 4 K. This signifies that YNi$_2$B$_2$C is not a simple BCS-like conventional superconductor. The poor data fitting shown in the left-side inset supports this argument. Significantly, similar observations have been observed in the two-gap superconductor MgB$_2$.[33,34]

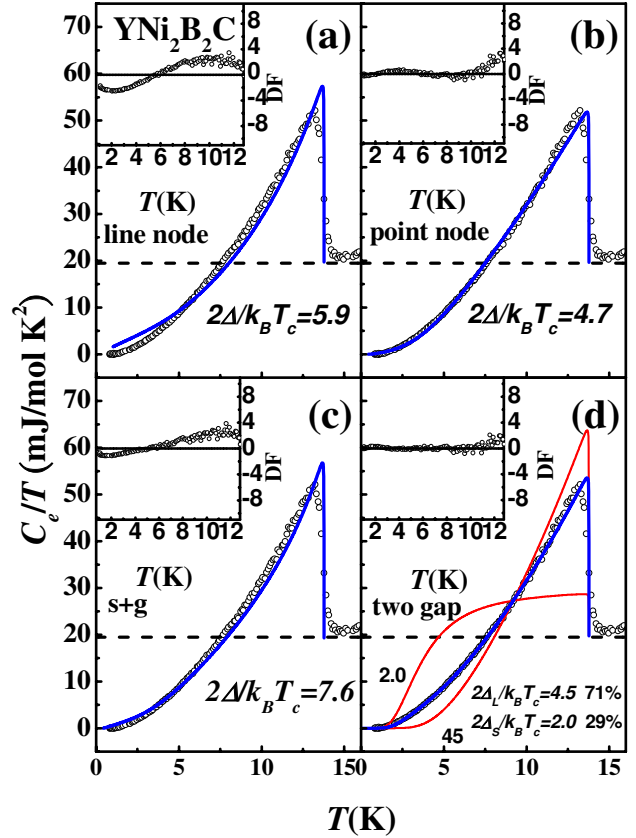

**FIG. 3.** Various fitting of $C_e/T$ vs. $T$ using (a) line-node, (b) point-node, (c) $s+g$ wave, and (d) two-gap models. The quality of each fitting is shown in the respective inset, where DF represents the difference between the calculated values and actual data.

To further explore the possibility of other unconventional superconducting order parameters in YNi$_2$B$_2$C, the data are fitted to line-node, point-node, $s+g$ wave and two-gap models, as shown in Figs. 3(a), (b), (c) and (d), respectively. The equation used in evaluating the superconducting-state electronic specific heat $C_e$ for these models is

$$C_e = 2N(0)\beta k \cdot \frac{1}{4\pi} \int_0^{2\pi} d\phi \int_0^{\pi} d\theta \sin\theta \int_{-\hbar\omega_D}^{\hbar\omega_D} -\frac{\partial f}{\partial E}(E^2 + \frac{1}{2}\beta \frac{d\Delta^2}{d\beta}) d\varepsilon, \quad (2)$$

where $N(0)$ is the density of states at the Fermi surface, $\beta=1/kT$, $E=(\varepsilon^2+\Delta^2)^{1/2}$, $f=(1+e^{\beta E})^{-1}$, $\Delta=\Delta_0$ the superconducting energy gap for isotropic $s$ wave, $\Delta=\Delta_0\cos n\phi$ for line nodes (to simplify, $n=2$ for $d$-wave), $\Delta=\Delta_0\sin n\theta$ for point nodes and $\Delta=\Delta_0/2(1-\sin^4\theta\cos 4\phi)$ for $s+g$ wave. For simplicity, the Fermi surface is assumed to be spherical except that, for the case of the line nodes, a two dimensional Fermi surface is applied. The deviation of the fit (DF) in the inset indicates clearly that the line-node model can not satisfactorily describe the data as shown in the Fig. 3(a). The fit of data to the point-node model is quite good at low temperatures, unless it becomes a little ambiguous at the high temperature region as shown in the Fig. 3(b). This can be related to the fact that $C_e$ in Eq. (2) is proportional to $T^3$ only at low temperatures in the point-node case, while the observed $T^3$-dependence of $C_e$ prevails almost throughout the entire superconducting state ($T < T_c$) as shown in the inset of Fig. 2 and also in several previous reports.[9,20,23] However, considering the experimental resolution, the point-node symmetry cannot be totally ruled out from the present specific-heat study. On the other hand, Eq. (2) gives $C_e \sim T^2$ at low temperatures for the $s+g$ wave case (as addressed in Ref. 26), thus the unacceptable fit is seen in the Fig. 3(c). Furthermore, from the analysis of the $H=8$ T data, the concentration of paramagnetic impurity is less than $10^{-3}$. Therefore, even considering the scenario of impurity scattering,[26] the observed $T^3$-dependence is difficult to reconcile with the $s+g$ wave model. Other than these three possibilities, the two-gap model, which has been successfully applied to $MgB_2$ superconductor,[34,35] was employed to further analyze the present data. In this two-gap scenario, two distinct gaps $2\Delta_L/k_BT_c=4.5$ and $2\Delta_S/k_BT_c=2.0$ with 71% and 29% in relative weight, respectively, are introduced. The excellent fit to the data is shown in Fig. 3(d). Consequently, the corresponding large gap $\Delta_L=2.67$ meV and the small gap $\Delta_S=1.19$ meV can be obtained. It is noted that the derived gap values are fairly consistent with the recent result from point contact spectroscopy,[30] where an anisotropic (or a multiband band) gap with $\Delta_{max}=2.4$ meV and $\Delta_{min}=1.5$ meV was revealed. Conceptually, the point-node and two-gap models are incompatible. However, for a highly anisotropic case it is not easy to experimentally distinguish between them.

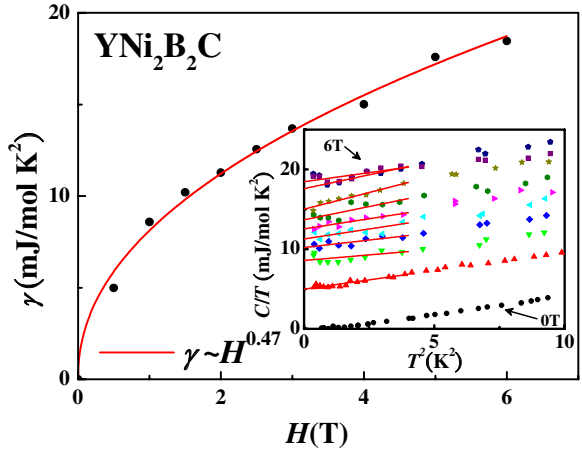

**FIG. 4.** Magnetic field dependence of electronic specific heat coefficient $\gamma(H)$ and the solid line representing $\gamma(H) \sim H^{0.47}$ is the best fit. The inset shows the determination of $\gamma(H)$ from the linear extrapolation of data for each field below 2 K in Fig. 1.

An alternative approach to study the superconducting order parameter is through the vortex excitations in the mixed state. Figure 4 shows the linear coefficient of electronic specific heat $\gamma(H)$ as a function of the applied magnetic fields. Each point was obtained from the linear extrapolation of data below 2 K for a given field as shown in the inset. The best fit yields $\gamma(H) \sim H^{0.47}$. Apparently, $\gamma(H)$ follows an $H^\alpha$ dependence with $\alpha$ being very close to 0.5 as reported previously.[9,16,20,21] Such a $\gamma(H) \sim H^{1/2}$ relation was used to support that $YNi_2B_2C$ is possibly a nodal superconductor[20-22] similar to a nodal $d$-wave cuprate superconductor[32,36] as predicted by Volovik.[37] However, the nonlinear $H$ dependence of $\gamma(H)$ has also been observed and successfully argued to be an intrinsic property of the two-gap superconductor $MgB_2$.[33,34,38,39]

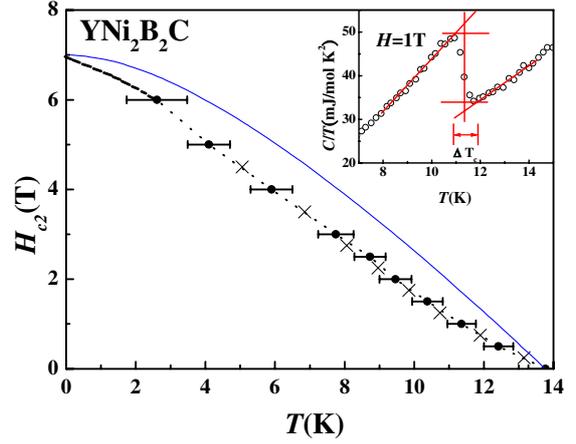

**FIG. 5.** Temperature dependence of upper critical field $H_{c2}(T)$ of $YNi_2B_2C$ obtained from Fig. 1. The crosses representing data from magnetization measurements provide a consistency check. The dashed line simply connects the data points, while the solid line is calculated from the WHH theory. Inset shows the determination of error bars.

Figure 5 shows the temperature dependence of the upper critical field $H_{c2}(T)$ of $YNi_2B_2C$ obtained from Fig. 1, where the solid line is based on the Werthamer-Helfand-Hohenberg (WHH) theory. The salient features are the positive curvature near $T_c$ and all $H_{c2}$ values being smaller than the theoretical prediction. Meanwhile, the $H_{c2}(T)$ behavior is consistent with that measured resistively on $YNi_2B_2C$ and $LuNi_2B_2C$,[28] from which Shulga et al. have successfully calculated such an $H_{c2}(T)$ behavior based on the two-band model of superconductivity.[28] Just as convincing, similar observations have also been proposed to reflect the intrinsic property of two-gap superconductivity in $MgB_2$.[40]

In summary, the magnetic-field dependence of low temperature specific heat of $YNi_2B_2C$ has been analyzed using various superconducting symmetry order parameters. Several critical findings include: (1) At zero magnetic field, the superconducting gap function of $YNi_2B_2C$ is better described by the point-node and two-gap models than other models based on single isotropic $s$ wave, line nodes or $s+g$ wave. (2) The two gap values, $\Delta_L=2.67$ meV and $\Delta_S=1.19$ meV obtained from the data fitting to the two-gap model are consistent with recent reported multiband gaps determined from point-contact spectroscopy. The observed nonlinear relation $\gamma(H) \sim H^{0.47}$ in the mixed state can be explained by either nodal or two-gap superconductivity. (3) The positive curvature observed in $H_{c2}(T)$ near $T_c$ is similar to that in $MgB_2$. These findings provide calorimetric evidence to support that $YNi_2B_2C$, might be a point-node or likely as $MgB_2$ a two-gap superconductor.

This work was supported by National Science Council of Republic of China under contract Nos. NSC93-2112-M110-001 and NSC93-2112-M009-015. The fruitful discussion with Professors W. P. Su, H. C. Ho and Dr. N. Nakai is appreciated.

*Corresponding author: yang@mail.phys.nsysu.edu.tw

[1] R. Nagarajan, C. Mazumdar, Z. Hossain, S. K. Dhar, K. V. Gopalakrishnan, L. C. Gupta, C. Godart, B. D. Padalia, R. Vijayaraghavan, Phys. Rev. Lett. **72**, 274 (1994).

[2] R. J. Cava, H. Takagi, H. W. Zandbergen, J. J. Krajewski, W. F. Peck, T. Siegrist, B. Batlogg, R. B. Vandover, R. J. Felder, K. Mizuhashi, J. O. Lee, H. Eisaki, S. Uchida, Nature (London) **367**, 252 (1994).

[3] K. –H. Muller and V. N. Narozhnyi, Rep. Prog. Phys. **64**, 943 (2001), and references cited therein.

[4] B. H. Brandow, Philos. Mag. **83**, 2487 (2003).

[5] L. F. Mattheiss, Phys. Rev. B **49**, R13279 (1994).

[6] S. A. Carter, B. Batlogg, R. J. Cava, J. J. Krajewski, W. F. Peck, Jr., and H. Takagi, Phys. Rev. B **50**, R4216 (1994).

[7] H. Michor, T. Holubar, C. Dusek, and G. Hilscher, Phys. Rev. B **52**, 16165 (1995).

[8] R. S. Gonnelli, A. Morello, G. A. Ummarino, V. A. Stepanov, G. Behr, G. Graw, S. V. Shulga, and S. L. Drechsler, Int. J. Mod. Phys. B **14**, 2840 (2000).

[9] M. Nohara, H. Suzuki, N. Mangkorntong, and H. Takagi, Physica C **341-348**, 2177 (2000).

[10] I. S. Yang, M. V. Klein, S. L. Cooper, P. C. Canfield, B. K. Cho, and S. –I. Lee, Phys. Rev. B **62**, 1291 (2000).

[11] K. Izawa, A. Shibata, Y. Matsuda, Y. Kato, H. Takeya, K. Hirata, C. J. van der Beek, and M. Konczykowski, Phys. Rev. Lett. **86**, 1327 (2001).

[12] K. Izawa, K. Kamata, Y. Nakajima, Y. Matsuda, T. Watanabe, M. Nohara, H. Takagi, P. Thalmeier, and K. Maki, Phys. Rev. Lett. **89**, 137006 (2002).

[13] E. Boaknin, R. W. Hill, C. Proust, C. Lupien, L. Taillefer, and P. C. Canfield, Phys. Rev. Lett. **87**, 237001 (2001).

[14] Y. Matsuda and K. Izawa, Physica C **388-389**, 487 (2003).

[15] T. Park, M. B. Salamon, E. M. Choi, H. J. Kim, and S. I. Lee, Phys. Rev. Lett. **90**, 177001 (2003).

[16] T. Park, E. E. M. Chia, M. B. Salamon, E. Bauer, I. Vekhter, J. D. Thompson, E. M. Choi, H. J. Kim, S. I. Lee, P. C. Canfield, Phys. Rev. Lett. **92**, 237002 (2004).

[17] P. Raychaudhuri, D. Jaiswal-Nagar, G. Sheet, S. Ramakrishnan, and H. Takeya, Phys. Rev. Lett. **93**, 156802 (2004).

[18] T. Watanabe, M. Nohara, T. Hanaguri, and H. Takagi, Phys. Rev. Lett. **92**, 147002 (2004).

[19] H. Nishimori, K. Uchiyama, S. Kaneko, A, Tokura, H. Takeya, K. Hirata, and N. Nishida, J. Phys. Soc. Jpn. **73**, 3247 (2004).

[20] M. Nohara, M. Isshiki, H. Takagi, and R. J. Cava, J. Phys. Soc. Jpn. **66**, 1888 (1997).

[21] M. Nohara, M. Isshiki, F. Sakai, and H. Takagi, J. Phys. Soc. Jpn. **68**, 1078 (1999).

[22] G. Wang and K. Maki, Phys. Rev. B **58**, 6493 (1998).

[23] K. J. Song, C. Park, S. S. Oh, Y. K. Kwon, J. R. Thompson, D. G. Mandrus, D. McK. Paul, and C. V. Tomy, Physica C **398**, 107 (2003).

[24] K. Maki, P. Thalmeier, and H. Won, Phys. Rev. B **65**, 140502(R) (2002).

[25] Q. Yuan and P. Thalmeier, Phys. Rev. B **68**, 174501 (2003).

[26] Q. Yuan, H. Y. Chen, H. Won, S. Lee, K. Maki, P. Thalmeier, and C. S. Ting, Phys. Rev. B **68**, 174510 (2003).

[27] H. Doh, M. Sigrist, B. K. Cho, and S. –I. Lee, Phys. Rev. Lett. **83**, 5350 (1999)

[28] S. V. Shulga, S. –L. Drechsler, G. Fuchs, K. –H. Muller, K. Winzer, M. Heinecke, and K. Krug, Phys. Rev. Lett. **80**, 1730 (1998).

[29] S. Muhhopadhyay, G. Sheet, P. Raychaudhuri, and H. Takeya, Phys. Rev. B **72**, 014545 (2005).

[30] D. L. Bashlakov, Y. G. Naidyuk, I. K. Yanson, S. G. Wimbush, B. Holzapfel, G. Fuch, and S. L. Drechsler, Supercond. Sci. Technol. **18**, 1094 (2005).

[31] B. K. Cho, P. C. Canfield, L. L. Miller, D. C. Johnston, W. P. Beyermann, and A. Yatskar, Phys. Rev. B **52**, 3684 (1995).

[32] S. J. Chen, C. F. Chang, H. L. Tsay, H. D. Yang, and J. –Y. Lin, Phys. Rev. B **58**, R14753 (1998).

[33] H. D. Yang, J. –Y. Lin, H. H. Li, F. H. Hsu, C. J. Liu, S. –C. Li, R. –C. Yu, and C. –Q. Jin, Phys. Rev. Lett. **87**, 167003 (2001)

[34] F. Bouquet, R. A. Fisher, N. E. Phillips, D. G. Hinks, and J. D. Jorgensen, Phys. Rev. Lett. **87**, 47001 (2001)

[35] F. Bouquet, Y. Wang, R. A. Fisher, D. G. Hinks, J. D. Jorgensen, A. Junod, and N. E. Phillips Europhys. Lett. **56**, 856 (2001).

[36] K. A. Moler, D. L. Sisson, J. S. Urbach, M. R. Beasley, A. Kapitulnik, D. J. Baar, R. Liang, and W. N. Hardy, Phys. Rev. B **55**, 3954 (1997).

[37] G. E. Volovik, JETP Lett. **58**, 469 (1993).

[38] F. Bouquet, Y. Wang, I. Sheikin, T. Plackowski, A. Junod, S. Lee, and S. Tajima, Phys. Rev. Lett. **89**, 257001 (2002).

[39] N. Nakai, M. Ichioka, and K. Machida, J. Phys. Soc. Jpn. **71**, 23 (2002).

[40] L. Lyard, P. Samuely, P. Szabo, T. Klein, C. Marcenat, L. Paulius, K. H. P. Kim, C. U. Jung, H. –S. Lee, B. Kang, S. Choi, S. –I. Lee, J. Marcus, S. Blanchard, A. G. M. Jansen, U. Welp, G. Karapetrov, and W. K. Kwok, Phys. Rev. B **66**, 180502(R) (2002).